 \documentclass[final,5p,times,twocolumn]{elsarticle}

\usepackage{amsmath}
\usepackage{amssymb}
\usepackage{lipsum}
\usepackage{float}
\usepackage{orcidlink}
\usepackage{textgreek}
\usepackage[mathlines]{lineno}

\journal{Physics Letters B}

\begin{document}

\begin{frontmatter}



\title{High-precision measurement of the kaonic hydrogen 1s level shift and width with SIDDHARTA-2}


\author[lnf]{M. Bazzi\fnref{eqAut}\fnref{eqAut}\orcidlink{0000-0002-1699-7138}}
\author[lnf,roma2]{F. Clozza\fnref{eqAut}\orcidlink{0009-0002-3298-0624}\corref{cor1}}
\cortext[cor1]{Corresponding author}
\ead{francesco.clozza@lnf.infn.it}
\author[lnf]{C. Guaraldo\fnref{eqAut,dec}\orcidlink{0000-0002-8923-3438}}
\author[lnf]{M. A. Iliescu\fnref{eqAut}\orcidlink{0009-0003-3859-5679}}
\author[lnf]{A. Scordo\fnref{eqAut}\orcidlink{0000-0002-7703-7050}}
\author[lnf]{F. Sgaramella\fnref{eqAut}\orcidlink{0000-0002-0011-8864}}
\author[lnf,cref,rom]{D. Sirghi\fnref{eqAut}\orcidlink{0009-0002-7486-025X}}
\author[lnf,rom]{F. Sirghi\fnref{eqAut}\orcidlink{0000-0002-6143-3200}}
\author[smi]{J. Zmeskal\fnref{eqAut,dec}\orcidlink{0000-0003-0815-0639}}
\author[pal,lnf]{L. Abbene\orcidlink{0000-0001-9633-6606}}
\author[smi]{C. Amsler\orcidlink{0000-0002-7695-501X}\fnref{ref1}}
\author[lnf,roma3]{F. Artibani\orcidlink{0009-0000-8905-3165}}
\author[polimi,infnMi]{G. Borghi\orcidlink{0000-0001-8488-4728}}
\author[zag,lnf]{D. Bosnar\orcidlink{0000-0003-4784-393X}}
\author[rom,lnf]{M. Bragadireanu\orcidlink{0009-0001-5217-6003}}
\author[pal,lnf]{A. Buttacavoli\orcidlink{0000-0002-7188-3651}}
\author[polimi,infnMi]{M. Carminati\orcidlink{0000-0001-9734-3007}}
\author[lnf]{A. Clozza\orcidlink{0000-0003-2133-1725}}
\author[lnf]{L. De Paolis\orcidlink{0000-0002-4203-9902}}
\author[prag,lnf]{R. Del Grande\orcidlink{0000-0002-7599-2716}}
\author[jag1,jag2,lnf]{K. Dulski\orcidlink{0000-0002-4093-8162}}
\author[polimi,infnMi]{C. Fiorini\orcidlink{0000-0002-1157-0143}}
\author[zag]{I. Fri\v{s}\v{c}i\'c\orcidlink{0000-0002-4743-0572}}
\author[rik]{M. Iwasaki\orcidlink{0000-0002-3460-9469}}
\author[jag1,jag2,lnf]{A. Khreptak\orcidlink{0000-0002-9482-9770}}
\author[lnf]{S. Manti\orcidlink{0000-0003-3770-0863}}
\author[smi]{J. Marton\orcidlink{0009-0003-1912-285X}\fnref{ref2}}
\author[lnf]{C. Milardi\orcidlink{0000-0002-6774-2848}}
\author[jag1,jag2]{P. Moskal\orcidlink{0000-0001-5644-5963}}
\author[uniper,infnPer,lnf]{F. Napolitano\orcidlink{0000-0002-8686-5923}}
\author[sendai]{H. Ohnishi\orcidlink{0000-0001-9427-1984}}
\author[cref,lnf]{K. Piscicchia\orcidlink{0000-0001-6879-452X}}
\author[pal,lnf]{F. Principato\orcidlink{0000-0003-2787-0877}}
\author[jag1]{M. Silarski\orcidlink{0000-0003-2206-0963}}
\author[jag1,jag2,lnf]{M. Skurzok\orcidlink{0000-0002-4794-5154}}
\author[lnf]{A. Spallone\orcidlink{0009-0000-2111-8014}}
\author[sendai,lnf]{K. Toho\orcidlink{0009-0001-3245-1418}}
\author[pal]{M. T\"uchler\orcidlink{0000-0003-0819-9870}}
\author[lnf]{O. Vazquez Doce\orcidlink{0000-0001-6459-8134}}
\author[lnf]{C. Curceanu\orcidlink{0000-0002-1990-0127}}

\fntext[eqAut]{This author contributed equally.}
\fntext[dec]{Deceased.}
\fntext[ref1]{Now at Marietta Blau Institute for Particle Physics, Vienna, Austria.}
\fntext[ref2]{Now at Atominstitut, Technische Universität Wien, Vienna, Austria.}

\affiliation[lnf]{organization={Laboratori Nazionali di Frascati INFN}, city={Frascati}, country={Italy}}
\affiliation[roma2]{organization={Università degli studi di Roma Tor Vergata, Dipartimento di Fisica}, city={Roma}, country={Italy}}
\affiliation[cref]{organization={Centro Ricerche Enrico Fermi - Museo Storico della Fisica e Centro Studi e Ricerche "Enrico Fermi"}, city={Roma}, country={Italy}}
\affiliation[pal]{organization={Department of Physics and Chemistry (DiFC) - Emilio Segrè, University of Palermo}, city={Palermo}, country={Italy}}
\affiliation[smi]{organization={Stefan Meyer Institute for Subatomic Physics}, city={Vienna}, country={Austria}}
\affiliation[roma3]{organization={Università degli studi di Roma Tre, Dipartimento di Fisica}, city={Roma}, country={Italy}}
\affiliation[polimi]{organization={Politecnico di Milano, Dipartimento di Elettronica, Informazione e Bioingegneria}, city={Milano}, country={Italy}}
\affiliation[infnMi]{organization={INFN Sezione di Milano}, city={Milano}, country={Italy}}
\affiliation[zag]{organization={Department of Physics, Faculty of Science, University of Zagreb}, city={Zagreb}, country={Croatia}}
\affiliation[rom]{organization={Horia Hulubei National Institute of Physics and Nuclear Engineering (IFIN-HH)}, city={Măgurele}, country={Romania}}
\affiliation[prag]{organization={Faculty of Nuclear Sciences and Physical Engineering, Czech Technical University in Prague}, city={Prague}, country={Czech Republic}}
\affiliation[jag1]{organization={Faculty of Physics, Astronomy, and Applied Computer Science, Jagiellonian University}, city={Kraków}, country={Poland}}
\affiliation[jag2]{organization={Center for Theranostics, Jagiellonian University}, city={Krakow}, country={Poland}}
\affiliation[rik]{organization={RIKEN}, city={Tokyo}, country={Japan}}
\affiliation[uniper]{organization={Dipartimento di Fisica e Geologia, Università degli studi di Perugia}, city={Perugia}, country={Italy}}
\affiliation[infnPer]{organization={INFN Sezione di Perugia}, city={Perugia}, country={Italy}}
\affiliation[sendai]{organization={Research Center for Accelerator and Radioisotope Science (RARIS), Tohoku University}, city={Sendai}, country={Japan}}

\begin{abstract}
Kaonic atoms provide a unique experimental probe of strong interaction in the low-energy regime. In particular, the strong-interaction-induced shift ($\varepsilon_{1\text{s}}$) and width ($\Gamma_{1\text{s}}$) of kaonic hydrogen directly constrain the low-energy antikaon-nucleon ($\bar{K}N$) interaction at threshold and the theoretical description of the $\Lambda$(1405) resonance. We report a new high-precision measurement of kaonic hydrogen X-ray transitions performed by the SIDDHARTA-2 experiment at the DA$\Phi$NE collider (INFN-LNF), based on an integrated luminosity of  237~pb$^{-1}$. The extracted values, $\varepsilon_{1\text{s}}$\,=\,$-$303.0\,$\pm$\,17.0\,(stat.)\,$\pm$\,2.5\,(syst.)~eV and $\Gamma_{1\text{s}}$\,=\,607\,$\pm$\,62\,(stat.)\,$\pm$\,6\,(syst.)~eV, represent the most precise determination to date, improving the precision by approximately a factor-of-two with respect to the previous SIDDHARTA measurement. These results significantly tighten the experimental constraints on theoretical description of the low-energy $\bar{K}N$ interaction.
\end{abstract}



\begin{keyword}
Kaonic hydrogen \sep X-ray spectroscopy \sep strong interaction \sep $\bar{K}N$ models



\end{keyword}

\end{frontmatter}




\section{Introduction}
Since their discovery in cosmic rays \cite{Rochester:1947mi}, kaons -- mesons containing a strange quark -- have been instrumental for the advance of particle and fundamental physics \cite{Aebischer:2025mwl}. In particular, kaonic atoms, atomic systems where an electron is replaced by a negatively charged kaon, provide a unique laboratory to probe the low-energy antikaon-nucleon ($\bar{K}N$) interaction and to experimentally access the strong interaction with strangeness in the low-energy regime \cite{Curceanu:2026zjg, Curceanu:2019uph}.
The strong interaction induces energy shifts ($\varepsilon$) and widths ($\Gamma$) of the lowest-$n$ atomic levels, which are extracted from high-precision measurements of the X-rays emitted during the atomic cascade towards these levels. In this framework, the shift and width of kaonic hydrogen 1s level, $\varepsilon_{1\text{s}}$ and $\Gamma_{1\text{s}}$, provide the most direct experimental access to the $\bar{K}N$ interaction at threshold ($\sim$\,1432~MeV) \cite{Shevchenko:2016wnu,Shevchenko:2025yxi}. Through the improved, summed-up Deser formula (Eq. \ref{eq:deser}) \cite{Shevchenko:2021swf}, they determine the complex $K^-p$ scattering length ($a_{K^-p}$) at threshold:
\begin{equation} \label{eq:deser}
    \varepsilon_{1\text{s}} + \frac{i}{2}\Gamma_{1\text{s}} = 2\alpha^3\mu^2a_{K^-p} \big/ \left[1+2\alpha\mu(\ln\alpha - 1)a_{K^-p}\right] \, ,
\end{equation}
where $\mu$ is the reduced mass of kaonic hydrogen and $\alpha$ the fine structure constant. 
The $K^-p$ scattering length relates to the isospin-dependent $\bar{K}N$ scattering lengths $a_{I=0,1}$ as:
\begin{equation}
    a_{K^-p} = \frac{1}{2}(a_0+a_1) \, .
\end{equation}

The low-energy $\bar{K}N$ interaction in the $I$\,=\,0 channel is strongly attractive and dynamically generates the $\Lambda(1405)$ below threshold \cite{Dalitz:1959dn, Hyodo:2011ur}. While its two-pole structure is well established theoretically \cite{Jido:2003cb, PDG:2019lam, Mai:2014xna}, the actual pole positions remain uncertain, making precise experimental constraints essential. The threshold scattering length $a_{K^-p}$, extracted from kaonic hydrogen, anchors the subthreshold $\bar{K}N$ amplitude and strongly impacts the properties of the $\Lambda(1405)$ and few- and many-body calculations.
Early measurements of kaonic hydrogen X-rays in the late 1970s and early 1980s \cite{Bird:1983yb, Izycki:1980uz, Davies:1979aj} yielded results in disagreement with theoretical expectations for $\varepsilon_{1\text{s}}$ and $\Gamma_{1\text{s}}$, a discrepancy known as the kaonic hydrogen puzzle \cite{Curceanu:2019uph}. This was later resolved by the KpX experiment at KEK-PS \cite{Iwasaki:1997wf}, which established the correct sign of the shift, and was subsequently confirmed by the DEAR experiment at DA$\Phi$NE \cite{DEAR:2005fdl}. To date, the most accurate measurement of kaonic hydrogen 1s shift and width was performed by the SIDDHARTA experiment \cite{SIDDHARTA:2011dsy} at the DA$\Phi$NE collider of the National Laboratories of Frascati (INFN-LNF) \cite{Milardi:2018sih,Milardi:2021khj,Zobov:2010zza}. This measurement has served as a key threshold input for phenomenological and SU(3) chirally motivated $\bar{K}N$ interaction potentials \cite{Shevchenko:2016wnu, Shevchenko:2025yxi}, providing essential constraints for the construction of three- and many-body models. 
Reducing the experimental uncertainty on $\varepsilon_{1\text{s}}$ and $\Gamma_{1\text{s}}$ therefore tightens the allowed parameter space of the models, sharpening their predictive capability. In particular, these improved constraints impact the description of the $\Lambda(1405)$ resonance, as a refined experimental input at threshold reduces the uncertainties on the position and width of the $\Lambda(1405)$ poles, as demonstrated in Ref. \cite{Mai:2020ltx}. These constraints also propagate to calculations of three- and many-body systems performed using Faddeev \cite{Faddeev:1961cu} and Faddeev-type AGS \cite{Alt:1967fx} equations, or variational techniques, which rely on the same $\bar{K}N$ interaction \cite{Shevchenko:2016wnu, Shevchenko:2025yxi}. Therefore, along with improving the description of the $\Lambda(1405)$, a more accurate determination of $\varepsilon_{1\text{s}}$ and $\Gamma_{1\text{s}}$, and thus of $a_{K^-p}$, narrows the theoretical spread of predicted binding energies and widths of the $K^-pp$ quasi-bound state and of the three-body calculations for kaonic deuterium \cite{Shevchenko:2025yxi, Revai:2014twa}, increasing the discriminatory power of forthcoming experimental searches \cite{J-PARCE15:2018zys, J-PARCE15:2020gbh}.

In this Letter we report a high-precision measurement of the strong-interaction-induced shift and width of the kaonic hydrogen 1s level performed by the SIDDHARTA-2 experiment. The accuracy achieved represents a significant improvement over previous best measurements, providing a new benchmark for constraining the low-energy $\bar{K}N$ interaction at threshold.

\section{Methods}
The SIDDHARTA-2 experiment at the DA$\Phi$NE $e^+e^-$ collider exploits low-momentum kaons from $\phi$ decays at rest to perform high-precision X-ray spectroscopy of kaonic atoms. Compared to the SIDDHARTA apparatus \cite{SIDDHARTA:2011dsy}, the experiment features an upgraded silicon drift detector (SDD) system \cite{Miliucci:2021wbj,Miliucci:2022lvn} with larger active area, improved timing resolution, and reduced electronic noise, resulting in significantly improved background rejection and energy resolution. A detailed description of the setup and its elements is given in Ref. \cite{Sirghi:2023wok}. These improvements provide a significantly enhanced experimental capability in the measurement of kaonic atoms X-ray transition lines, as demonstrated in Ref. \cite{Sgaramella:2024ehl}. Moreover, continued fine tuning of the DA$\Phi$NE collider reduced the machine background by a factor three compared to the SIDDHARTA experiment, while maintaining the same instantaneous luminosity \cite{Milardi:2024efr}. All these upgrades played a crucial role in enabling the improved precision measurement of kaonic hydrogen reported in this Letter.
The measurement was performed in two separate runs, from December 2023 to February 2024 and from April to May 2024, for a total integrated luminosity of 237~pb$^{-1}$. The cryogenic target cell was filled with gaseous hydrogen at 1.65\% of liquid-hydrogen density, maintained at 1.3~bar and 24~K. Collected data were processed to efficiently reduce the background and isolate X-ray transitions coming from kaonic hydrogen de-excitations measured by the SDDs system. The dominant background arises from electrons and positrons lost from the beams, which produce electromagnetic showers mainly via Touschek effect and beam-gas interactions \cite{Boscolo:2011zz}. These background events are suppressed by the kaon trigger system, which reduces them by nearly four orders of magnitude. Residual background from accidental coincidences induced by minimum ionizing particles is further suppressed using time-of-flight techniques to discriminate genuine $K^+K^-$ triggers from background events. A further selection is based on the timing capability of the SDDs system to select events correlated with the kaon trigger, yielding an additional factor-of-two reduction. More details on the data selection can be found in Refs. \cite{Sgaramella:2024ehl, Sgaramella:2024klx}. Dedicated calibration runs of the SDDs system were conducted using the X-ray K$_{\alpha}$ lines coming from Ti, Fe and Cu, as detailed in Ref. \cite{Sgaramella:2024ehl}. The systematic uncertainties on the energies of measured X-ray lines were determined from the energy residuals of these reference lines. Once the data were selected, a chi-squared fit of the final spectrum was performed covering an energy range from 4.2~keV to 14.5~keV to extract the energies of the observed X-ray transitions. The kaonic hydrogen lines were modeled by a set of Voigt profiles, sharing a common intrinsic width $\Gamma_{1\text{s}}$ and energy shift $\varepsilon_{1\text{s}}$, which jointly account for the strong-interaction effects on the 1s level. The transition energies were defined as $E = E^{\text{em}} + \varepsilon_{1\text{s}}$, where $E^{\text{em}}$, reported in Table \ref{tab:energies}, are fixed electromagnetic reference values computed with the Klein–Gordon equation including vacuum-polarization and recoil corrections \cite{Santos:2004bw,Karshenboim:2006zz,Karshenboim:2005am}, and $\varepsilon_{1\text{s}}$ is a free parameter of the fit. 
\begin{figure*}[t]
        \centering
        \includegraphics[width=0.70\textwidth]{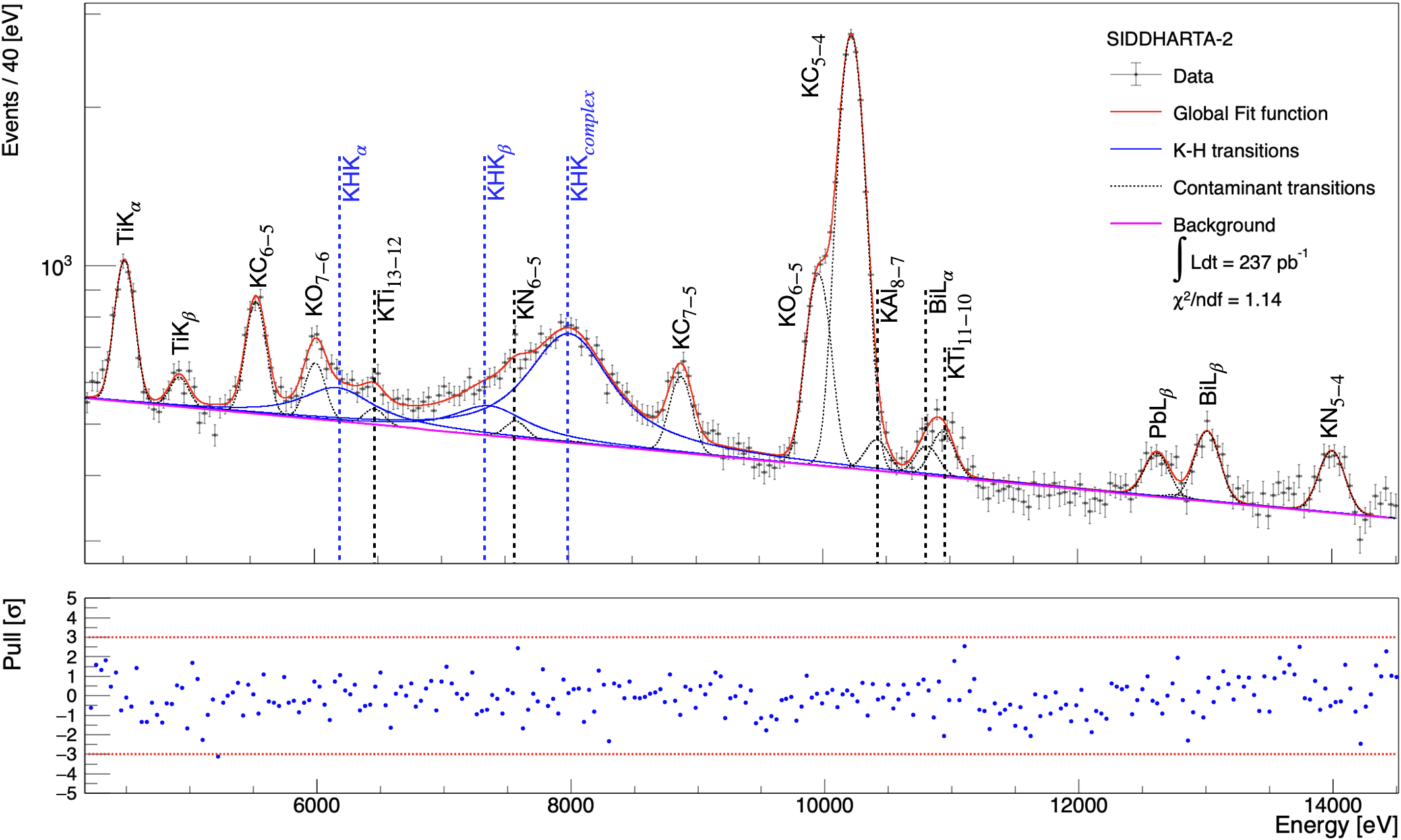}
        \caption{Fit of the kaonic hydrogen energy spectrum after event selection. Global fit (red), kaonic hydrogen lines (blue), residual contaminant lines (dashed black) and background (pink) are shown. In the bottom panel the pull plot is presented.}
        \label{fig:KH_spectrum}
\end{figure*}
Contaminant lines were described by Gaussian functions. Both Voigt and Gaussians profiles were defined with energy-dependent resolutions parametrized in terms of the Fano factor and electronic noise of the SDDs system \cite{Miliucci:2021wbj,perotti1999observed,Gysel:2003}, treated as free parameters of the fit. The residual background, primarily from DA$\Phi$NE electromagnetic events, was modeled by a decreasing exponential function summed to a constant offset. 

\begin{table}[H]
    \centering
    \small
    \caption{Electromagnetic transition energies for kaonic hydrogen X-ray lines \cite{Santos:2004bw,Karshenboim:2006zz,Karshenboim:2005am}.}
    \begin{tabular}{cc}
    \hline
    Transition & $E^{\text{em}}$ [eV] \\
    \hline
    $2\text{p} \rightarrow 1\text{s}$ & 6481.2 \\
    $3\text{d} \rightarrow 1\text{s}$ & 7678.0 \\
    $4\text{f} \rightarrow 1\text{s}$ & 8096.8 \\
    $5\text{g} \rightarrow 1\text{s}$ & 8290.6 \\
    $6\text{h} \rightarrow 1\text{s}$ & 8395.9 \\
    \hline
    \end{tabular}
    \label{tab:energies}
\end{table}

\begin{table*}[t]
    \centering
    \small
    \caption{Kaonic hydrogen strong-interaction shift $\varepsilon_{1\text{s}}$ and width $\Gamma_{1\text{s}}$ measured by SIDDHARTA-2, SIDDHARTA, KpX and DEAR. The corresponding $K^-p$ scattering lengths $a_{K^-p}$, computed with the improved summed-up Deser formula, are reported.}
    \begin{tabular}{cccc}
    \hline
    Experiment & $\varepsilon_{1\text{s}}$ [eV] & $\Gamma_{1\text{s}}$ [eV] & $a_{K^-p}$ [fm] \\
    \hline
    SIDDHARTA-2 (this work) & -303.0 $\pm$ 17.0 (stat.) $\pm$ 2.5 (syst.) & 607 $\pm$ 62 (stat.) $\pm$ 6 (syst.) & (-0.715$\pm$0.054) + $i$(0.905$\pm$0.091) \\
    SIDDHARTA \cite{SIDDHARTA:2011dsy} & -283 $\pm$ 36 (stat.) $\pm$ 6 (syst.) & 541 $\pm$ 89 (stat.) $\pm$ 22 (syst.) & (-0.68$\pm$0.11) + $i$(0.80$\pm$0.13) \\
    KpX \cite{Iwasaki:1997wf} & -323 $\pm$ 63 (stat.) $\pm$ 11 (syst.) & 407 $\pm$ 208 (stat.) $\pm$ 100 (syst.) & (-0.883$\pm$0.20) + $i$(0.62$\pm$0.35) \\
    DEAR \cite{DEAR:2005fdl} & -193 $\pm$ 37 (stat.) $\pm$ 6 (syst.) & 249 $\pm$ 111 (stat.) $\pm$ 30 (syst.) & (-0.49$\pm$0.10) + $i$(0.35$\pm$0.16) \\
    \hline
    \end{tabular}
    \label{tab:KHexp}
\end{table*}

\section{Results and discussion}
\begin{figure}[H]
        \centering
        \includegraphics[width=0.48\textwidth]{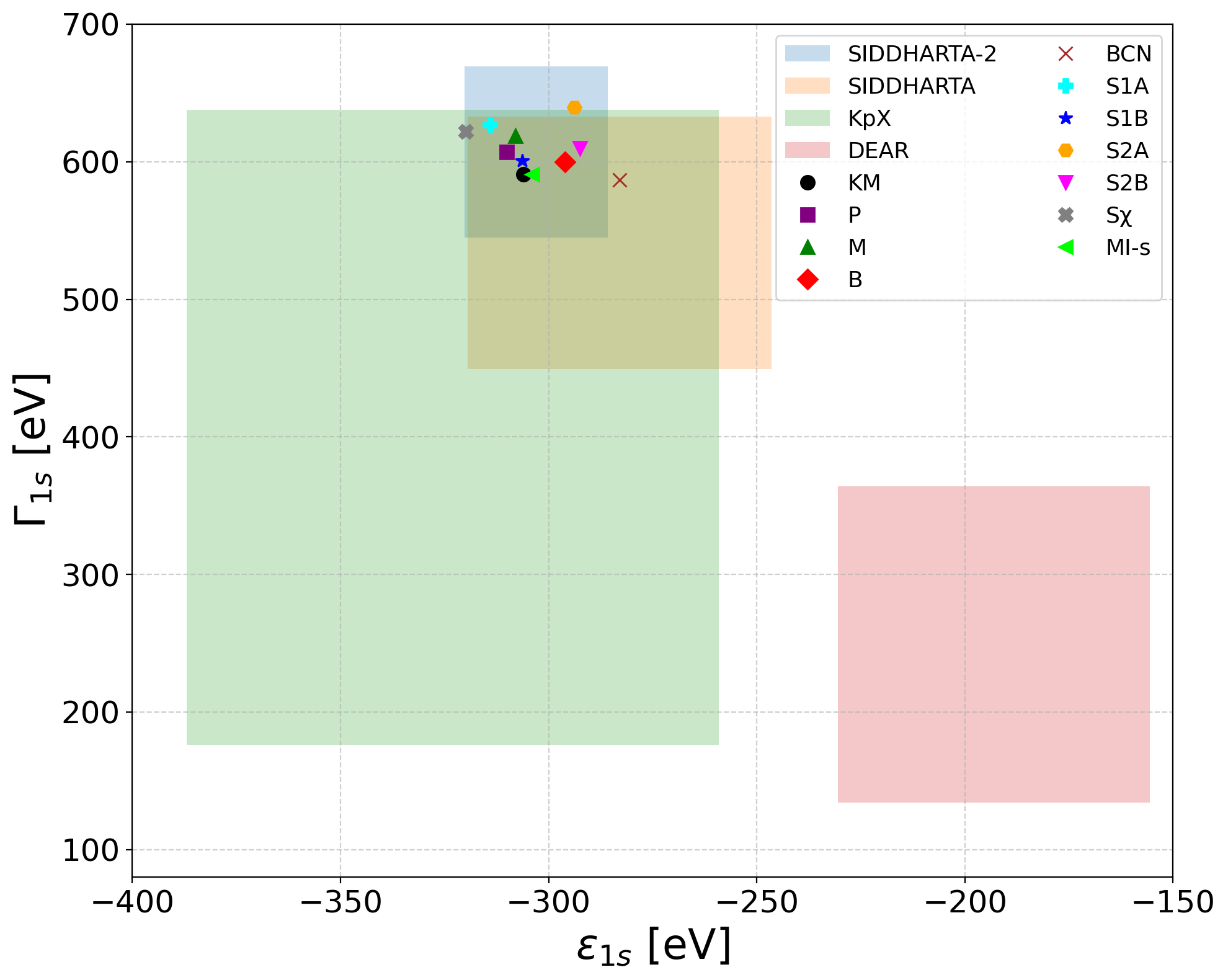}
        \caption{Experimental values of $\varepsilon_{1\text{s}}$ and $\Gamma_{1\text{s}}$ for kaonic hydrogen by SIDDHARTA-2 (this work, blue), SIDDHARTA (orange), KpX (green) and DEAR (red). The shaded areas represent the combined statistical and systematic uncertainties. Some of the most recent theoretical predictions for these quantities are reported. See text for the list of models.}
        \label{fig:shift_width}
\end{figure}

The fit of the kaonic hydrogen energy spectrum is shown in Fig. \ref{fig:KH_spectrum}. The X-ray K-series lines of kaonic hydrogen are observed, namely the 2p--1s (K$_{\alpha}$), 3d--1s (K$_{\beta}$) and K-complex, comprising unresolved transitions from higher-$n$ states (up to $n=6$) to the 1s level. Residual contaminant transitions, arising from accidental excitations of the setup materials (TiK$_{\alpha,\,\beta}$, BiL$_{\alpha,\,\beta}$, PbL$_{\beta}$) and from kaonic atoms formed in the solid elements of the apparatus (KC, KN, KO, KAl, KTi), are also observed. From the fit, the shift and width induced by the strong interaction on the kaonic hydrogen 1s level are determined to be $$\varepsilon_{1\text{s}} = -303.0 \pm 17.0\,(\text{stat}) \pm 2.5\,(\text{syst}) \; \text{eV}\,,$$ $$\Gamma_{1\text{s}} = 607 \pm 62\,(\text{stat}) \pm 6\,(\text{syst}) \; \text{eV}.$$ 
The systematic uncertainty on $\varepsilon_{1\text{s}}$ is dominated by the energy calibration while other contributions are negligible \cite{Sgaramella:2024klx}, and that on $\Gamma_{1\text{s}}$ arises mainly from uncertainties in the Fano factor and electronic noise, which determine the detectors' energy resolution. This result corresponds to approximately a factor-of-two improvement in precision with respect to SIDDHARTA. Fig. \ref{fig:shift_width} shows the $\varepsilon_{1\text{s}}$ and $\Gamma_{1\text{s}}$ experimental values of kaonic hydrogen measured by SIDDHARTA-2 in comparison to the previous ones of SIDDHARTA \cite{SIDDHARTA:2011dsy}, KpX \cite{Iwasaki:1997wf} and DEAR \cite{DEAR:2005fdl}.
For each experiment, the shaded area indicates the uncertainty region in the two-dimensional ($\varepsilon_{1\text{s}}$, $\Gamma_{1\text{s}}$) plane, centered on the measured values and obtained by combining statistical and systematic uncertainties in quadrature. Some of the most recent theoretical calculations of $\varepsilon_{1\text{s}}$ and $\Gamma_{1\text{s}}$ for kaonic hydrogen are also reported in the plot. We refer to these models as Kyoto-Munich (KM) \cite{Ikeda:2012au}, Prague (P) \cite{Cieply:2011nq}, Murcia (M) \cite{Guo:2012vv}, Bonn (B) \cite{Mai:2012dt}, Barcelona (BCN) \cite{Feijoo:2018den}, Shevchenko (S1A, S1B, S2A, S2B, S$\chi$) \cite{Shevchenko:2025yxi} and Mizutani (MI-s) \cite{Mizutani:2012gy}. 

The enhanced precision of the SIDDHARTA-2 measurement reduces the allowed region in the ($\varepsilon_{1\text{s}}$, $\Gamma_{1\text{s}}$) parameter space by approximately a factor-of-three with respect to SIDDHARTA, providing tighter constraints on $\bar{K}N$ theoretical models at threshold. All recent theoretical predictions remain compatible with the present measurement within uncertainties. Table \ref{tab:KHexp} summarizes the experimental values of the strong-interaction shift $\varepsilon_{1\text{s}}$ and width $\Gamma_{1\text{s}}$ of kaonic hydrogen measured by SIDDHARTA-2, SIDDHARTA \cite{SIDDHARTA:2011dsy}, KpX \cite{Iwasaki:1997wf} and DEAR \cite{DEAR:2005fdl}, together with the corresponding $K^-p$ scattering lengths at threshold derived using the improved summed-up Deser formula \cite{Shevchenko:2021swf}. The SIDDHARTA-2 result provides the most precise determination of both $\varepsilon_{1\text{s}}$ and $\Gamma_{1\text{s}}$ to date, leading to significantly reduced uncertainties on the real and imaginary parts of $a_{K^-p}$, and consequently on the isospin-dependent $\bar{K}N$ scattering lengths. This improved precision provides substantially more stringent experimental input for the two-body $\bar{K}N$ interaction, three- and many-body calculations and for constraining the pole structure of the $\Lambda(1405)$.

\section{Summary and conclusions}
We have presented a high-precision measurement of the strong-interaction-induced shift and width of the kaonic hydrogen 1s level performed by the SIDDHARTA-2 experiment at DA$\Phi$NE. The extracted values, $\varepsilon_{1\text{s}}=-303.0\pm17.0\,(\text{stat.})\pm2.5\,(\text{syst.})$~eV and $\Gamma_{1\text{s}}=607\pm62\,(\text{stat.})\pm6\,(\text{syst.})$~eV, constitute the most precise determination to date. The achieved accuracy significantly improves the experimental constraints on the low-energy $\bar{K}N$ interaction, leading to a more precise determination of both the complex $K^-p$ scattering length and the isospin-dependent $\bar{K}N$ scattering lengths, while also providing a refined input for approaches based on chiral dynamics and phenomenological potentials. These results have direct implications for the description of the $\Lambda(1405)$ resonance, whose properties are strongly linked to the subthreshold behavior of the $\bar{K}N$ amplitude, and for calculations of few- and many-body antikaonic systems, reducing the uncertainties in calculations relevant for kaonic deuterium and for quasi-bound states such as $K^-pp$. The measurement reported in this Letter establishes a new benchmark for experimental input to low-energy $\bar{K}N$ physics, enhancing the predictive power of theoretical models and supporting ongoing and future experimental studies in the strangeness sector.

\section*{Acknowledgements}
We thank C. Capoccia from LNF-INFN and H. Schneider, L. Stohwasser, and D. Pristauz-Telsnigg from Stefan Meyer-Institut for their fundamental contribution in designing and building the SIDDHARTA-2 setup. We thank as well the INFN, INFN-LNF and the DA$\Phi$NE staff for the excellent working conditions and permanent support.
Part of this work was supported by the Austrian Science Fund (FWF): [P24756-N20 and P33037-N]; the Croatian Science Foundation under the project IP-2022-10-3878; the EU STRONG-2020 project (Grant Agreement No. 824093); the EU Horizon 2020 project under the MSCA (Grant Agreement 754496); the Japan Society for the Promotion of Science JSPS KAKENHI Grant No. JP18H05402; the SciMat and qLife Priority Research Areas budget under the program Excellence Initiative - Research University at the Jagiellonian University, and the Polish National Agency for Academic Exchange (Grant No. PPN/BIT/2021/1/00037); the EU Horizon 2020 research and innovation programme under project OPSVIO (Grant Agreement No. 101038099). This work was also supported by the Italian Ministry for University and Research (MUR), under PRIN 2022 PNRR project CUP: B53D23024100001. This article/publication is based upon work from COST ActionCA24131-ENRICH, supported by COST (European Cooperation in Science and Technology, http://www.cost.eu/).




\bibliographystyle{elsarticle-num-names} 
\bibliography{ref}






\end{document}